\input harvmac.tex 

\def\Asym#1#2{\vcenter{\vbox{\drawbox{#1}{#2}
              \kern-#2pt       
              \drawbox{#1}{#2}}}}

\Title{\vbox{\rightline{hep-th/9803060} \rightline{CERN-TH/98-68}
}}
{\vbox{\centerline{N=1,2 4D Superconformal Field Theories}
\centerline{and Supergravity in $AdS_5$}}}


\centerline{Sergio Ferrara} \smallskip{\it
\centerline{CERN Geneva, Switzerland}}
\centerline{\tt
Sergio.Ferrara@cern.ch}

\centerline{Alberto Zaffaroni} \smallskip{\it
\centerline{CERN Geneva, Switzerland}}
\centerline{\tt
Alberto.Zaffaroni@cern.ch}

\vskip .1in


\noindent
We consider D3 branes world-volume theories sustaining $N=1,2$ superconformal field
theories. Under the assumption that these theories
are dual to $N=2,4$ supergravities in $AdS_5$, we explore the general structure
of the latter and discuss some issues when comparing the bulk theory to the boundary singleton theory.
\vskip 4truecm
\noindent
CERN-TH/98-68
\Date{March 98}

\lref\warner{M. Gunaydin, L. J. Romans and N. P. Warner, Nucl. Phys. B272 (1986) 598.}
\lref\malda{ J. M. Maldacena, {\it The Large N Limit of Superconformal Field Theories and Supergravity},  hep-th/9705104.}
\lref\maldatwo{ N. Itzhaki, J. M. Maldacena, J. Sonnenschein and S. Yankielowicz, {\it Supergravity and The Large N Limit of Theories With Sixteen Supercharges}, hep-th/9802042.}
\lref\witten{E. Witten, {\it Anti-de Sitter Space And Holography}, hep-th/9802150.}
\lref\romans{L. J. Romans, Nucl. Phys. B267 (1986) 433.}
\lref\pol{S. S. Gubser, I. R. Klebanov and A. M. Polyakov, {\it Gauge Theory Correlators from Non-Critical String Theory}, hep-th/9802109.}
\lref\fer{S. Ferrara and  C. Fronsdal, {\it  Conformal Maxwell theory as a singleton field theory on $ADS_5$, IIB three branes
     and duality}, hep-th/971223.}
\lref\fz{S. Ferrara and B. Zumino, Nucl. Phys. B87 (1975) 207.}
\lref\fertwo{S. Ferrara and  C. Fronsdal, {\it Gauge Fields as Composite Boundary Excitations}, hep-th/9802126.}
\lref\van{H. J. Kim, L. J. Romans and P. van Nieuwenhuizen, {\it The Mass Spectrum Of Chiral N=2 D=10 supergravity on $S^5$}, Phys. Rev. D23 (1981) 1278.}
\lref\roo{E. Bergshoeff, M. De Roo and B de Wit, {\it Extended Conformal Supergravity}, Nucl. Phys. B182 (1981) 173.}
\lref\stelle{P. Howe, K. S. Stelle and P. K. Townsend, {\it Supercurrents}, Nucl. Phys. B192 (1981) 332.}
\lref\kleb{C. W. Gibbons and P. K. Townsend, Phys. Rev. Lett. 71 (1993) 3754; M. P. Blencowe and M. J. Duff, Phis. Lett. B203 (1988) 229; Nucl. Phys B310 (1988), 389; M. J. Duff, Class. Quantum Grav. 5 (1988) 189; E. Bergshoeff, M. J. Duff, C. N. Pope and E. Sezgin, Phys. Lett. B199 (1988) 69; H. Nicolai, E. Sezgin and Y. Tanii, Nucl. Phys B305 (1988) 483.}
\lref\awada{M. Awada and P. K. Townsend, Nucl. Phys. B255 (1985) 617.}
\lref\sken{K. Sfetsos and K. Skenderis, {\it Microscopic derivation of the Bekenstein-Hawking entropy formula for
     non-extremal black holes}, hep-th/9711138.}
\lref\skentwo{H. J. Boonstra, B. Peeters and K. Skenderis, {\it Branes and anti-de Sitter spacetimes}, hep-th/9801076.}
\lref\kall{P. Claus, R. Kallosh and A. Van Proeyen, {\it M 5-brane and superconformal (0,2) tensor multiplet in 6 dimensions}, hep-th/9711161.}
\lref\kalltwo{ R. Kallosh, J. Kumar and  A. Rajaraman, {\it Special Conformal Symmetry of Worldvolume Actions}, hep-th/9712073.}
\lref\kallthree{P. Claus, R. Kallosh, J. Kumar, P. Townsend and A. Van Proeyen, {\it Conformal theory of M2, D3, M5 and D1+D5 branes}, hep-th/9801206.}
\lref\ooguri{G. T. Horowitz and H. Ooguri, {\it Spectrum of Large N Gauge Theory from Supergravity}, hep-th/9802116.}
\lref\gun{M. Gunaydin and D. Minic {\it Singletons, Doubletons and M-theory}, hep-th/9802047}
\lref\silv{S. Kachru and  E. Silverstein, {\it 4d Conformal Field Theories and Strings on Orbifolds}, hep-th/9802183.}
\lref\fztwo{S. Ferrara and B. Zumino, Nucl. Phys. B134 (1978) 301.}
\lref\guntwo{M. Gunaydin, L. J. Romans and N. P. Warner, Phys. Lett. 154B (1985) 268; M. Pernici, K. Pilch and P. van Nieuwenhuizen, Nucl. Phys. B259 (1985) 460.}
\lref\pilch{M. Pernici, K. Pilch and P. van Nieuwenhuizen, Phys. Lett. 143B (1984) 103.}
\lref\fr{M. Flato and C. Fronsdal, J. Math. Phys. 22 (1981) 1100; Phys. Lett. B172 (1986) 412.}
\lref\frtwo{M. Flato and C. Fronsdal, Lett. Math. Phys 2 (1978) 421; Phys. Lett. 97B (1980) 236.}
\lref\frthree{E. Angelopoulos, M. Flato, C. Fronsdal and D. Sternheimer, Phys. Rev. D23 (1981) 1278.}
\lref\sjr{S. J. Rey and  J. Yee, {\it Macroscopic Strings as Heavy Quarks of Large N Gauge Theory and Anti-de Sitter
     Supergravity}, hep-th/9803001.}
\lref\maldathree{J. M. Maldacena, {\it  Wilson loops in large N field theories}, hep-th/9803002.}
\lref\arefa{ I. Ya. Aref'eva and I. V. Volovich, {\it , Field Theories in Anti-De Sitter Space and Singletons}, hep-th/9803028.}
\lref\oz{O. Aharony, Y. Oz and  Z. Yin, {\it M Theory on $AdS_p\times S^{11-p}$ and Superconformal Field Theories}, hep-th/9803051.}
\lref\min{S. Minwalla, {\it Particles on $AdS_{4/7}$ and Primary Operators on $M_{2/5}$ Brane Worldvolumes}, hep-th/9803053.}
\lref\sierra{M. Gunaydin, G. Sierra and P. K. Townsend, Nucl. Phys. B253 (1985) 573.}
\lref\torin{L. Castellani, A. Ceresole, R. D'Auria, S. Ferrara, P. Fr\'e and M. Trigiante, {\it G/H M-branes and $AdS_{p+2}$ Geometries}, hep-th/9803039.}
\lref\ferthree{S. Ferrara, C, Fronsdal and A. Zaffaroni,{\it On $N=8$ Supergravity in $AdS_5$ and $N=4$ Superconformal Yang-Mills theory.}, hepth/9802203.}
\newsec{Introduction.}
Recently, a remarkable correspondence between d-dimensional conformal
field theories and supergravity in $AdS_{d+1}$ has been explored \refs{\malda,\fer,\maldatwo,\ooguri,\pol,\fertwo,\witten,\ferthree,\sjr,\maldathree,\arefa,\oz,\min}\foot{The close connection between the $AdS_{p+2}$ geometry and the p-brane dynamics has been also investigated in a series of papers \refs{\kleb,\sken,\skentwo,\kall,\kalltwo,\kalltwo,\kallthree,\gun}.}.

\lref\ein{M. J. Duff, H. Lu, C. W. Pope and E. Sezgin, Phys. Lett B371 (1996) 206, hep-th/9511162.}

The original duality of ref. \malda, between p-brane dynamics and supergravity
in its nearly horizon geometry, typically requiring a $AdS_{p+2}\times {\cal M}_{d-p-2}$ background\foot{A way of obtaining branes with lower supersymmetry is by identifying ${\cal M}_{d-p-2}$ with a suitable Einstein space as discussed in \refs{\ein,\torin}.}, has been further confirmed in identifying massless
excitations in the bulk with {\it singletons} composite operators on the
anti-De-Sitter boundary \refs{\fer,\fertwo} and further extending this relation by postulating the recipe that, in some suitable limit (of the parameter space), the generating functional for the boundary correlators of singleton composite fields is reproduced by the $AdS_{p+2}$ supergravity action \refs{\pol,\witten}. This analysis has been done for the $p=3$ case when 4d superconformal field
theories are better known.

\lref\silver{ S. Kachru and E. Silverstein, {\it 4d Conformal Field Theories and Strings on Orbifolds}, hep-th/9802183.}
\lref\berkooz{M. Berkooz, {\it A Supergravity Dual of a (1,0) Field Theory in Six Dimensions}, hep-th/9802195.}
\lref\vafa{A. Lawrence, N. Nekrasov and C. Vafa {\it On Conformal Field Theories in Four Dimensions}, hep-th/9803015.}

This recipe has been further extended to include full supersymmetry in the
maximal symmetric case, corresponding to ${\cal M}_{d-p-2}\times S_{d-p-2}$
($S_5$ for $d=10,p=3$), in \refs{\ferthree,\oz,\min} and models with lower supersymmetry in the world-volume theory started to be explored \refs{\silver,\berkooz,\vafa}.

If the correspondence is at work, for any given $N=1,2$ superconformal field theory in four dimensions, there should be a corresponding supergravity
theory in $AdS_5$. To gain new insights on the knowledge
of the latter is the purpose of the present paper.

One of the important point about five-dimensional supergravity theories is that
they can be regarded as Chern-Simons theories of a {\it gauge} symmetry $G$, which
is identified with the {\it rigid} symmetry on the world-volume theory.

The rigid symmetry is typically $G=U(2)\times G^{\prime}$ for $N=2$ and
$U(1)\times G^{\prime}$ in $N=1$ theories in which $U(2)$ and $U(1)$ are the R-symmetries which are embedded in the $U(2,2/2)$ and $U(2,2/1)$ superconformal algebras.

As pointed out in \witten, the Chern-Simons coupling in the 5d theory just reproduces the anomaly of the 4d boundary theory. This, of course, produces a simple and unique result in the $N=4$ super Yang-Mills case, in which $G=SU(4)$.
However, in $N=1$ and $N=2$ theories, the fact that the world-volume anomalies of the global
symmetries must match the Chern-Simons coupling in the 5d theory, gives a
rather remarkable test and in fact predicts to a large extent the structure of the bulk theory, which is the gauged $N=2$ and $N=4$ supergravity on $AdS_5$.

The paper is organized as follows. In section 2, we give a complete classification of the relevant multiplets for the $U(2,2/N)$ superalgebras ($N=1,2$). They
fall in three categories, as for the $N=4$ case, namely, singletons, massless
and massive multiplets.
In section 3, we recall some properties of the basic $N=2$ and $N=4$ supergravities in $AdS_5$, in particular the structure of their Chern-Simons terms. In
section 4, we compare world-volume superconformal field theories with their {\it dual} supergravity in $AdS_5$. The paper ends with a paragraph of conclusions and outlook.

\newsec{Unitary irreducible representations of $U(2,2/1)$ and $U(2,2/2)$.}
In the present paper, we will construct the relevant representations of the
$N=1$ and $N=2$ superconformal algebras, in their realization on the boundary singleton field theory (on ${\cal M}_4$) and in the bulk supergravity theory (on $AdS_5$). The main ingredient will be the fact that {\it massless fields}
in $AdS_5$ correspond to composites of singletons fields on the boundary, while singletons fields are just massless conformal fields on the boundary.

For the bosonic subgroup $SU(2,2)$ (the conformal group) of the superalgebra,
this analysis was carried out in \refs{\fer,\fertwo}, and its $N=4$ extension
was discussed in \ferthree. 

It is the aim of this section first to extend the singleton representation of $SU(2,2)$, namely $D(2,1,0)+D(3/2,1/2,0)+D(1,0,0)$ to the supersingleton, then to
extend the massless representations  of $SU(2,2)$ in $AdS_5$ to the corresponding
supermultiplets.

\lref\list{P. Howe, K. Stelle and P. West, Phys. Lett. 124B (1983) 55; 
F. X. Dong, T. S. Tu, P. Y. Xue and X. J. Zhou, Phys. Lett 140B (1984) 333;
I. G. Koh and S. Rajpoot, Phys. Lett. 135B (1984); J. P. Derendinger, S. Ferrara and A. Masiero, Phys. Lett 143B (1984) 133;
A. Parkes and P. West, Phys. Lett 138B (1984) 99; P. West, Phys. Lett. 137B (1984) 371; D. R. T. Jones and L. Mezincescu, Phys. Lett. 138B (1984) 293; S. Hamidi, J. Patera and J. Schwarz, Phys. Lett. 141B (1984) 349; S. Hamidi and J. Schwarz, Phys. Lett. 147B (1984) 301; W. Lucha and H. Neufeld, Phys. Lett. 174B
(1986) 186, Phys. Rev. D34 (1986) 1089; D. R. T Jones, Nucl. Phys. B277 (1986) 153; A. V. Ermushev, D. I. Kazakov and O.V. Tarasov, Nucl. Phys. B281 (1987) 72; X. D. Jiang and X. J. Zhou, Phys. Rev. D42 (1990) 2109; D. I. Kazakov, Mod. Phys. Lett. A2 (1987) 663; O. Piguet and K. Sibold, Int. J. Mod. Phys. A1 (1986)
913; Phys. Lett 177B (1986) 373; C. Lucchesi, O. Piguet and K. Sibold, {\it Conf. on Differential Geometrical Methods in Theoretical Physics}, Como 1987, Helv.
Phys. Acta 61 (1988) 321; R. G. Leigh and M. J. Strassler, Nucl. Phys. B447 (199, hep-th/9503121.}

Let us first consider the $N=2$ case. The $N=2$ superconformal algebra has a
$SU(2)\times U(1)$ R-symmetry. If the $N=2$ superconformal singleton theory has no other flavour symmetry then the corresponding $N=4$ supergravity
theory will be the gauged $SU(2)\times U(1)$ supergravity, with possible additional matter multiplets. If on the other hand, the superconformal theory has an additional flavour symmetry $G^{\prime}$,
then the corresponding supergravity theory will have certain
additional vector multiplets, which are the Yang-Mills multiplet of $G^{\prime}$. All these multiplets will arise as composite boundary excitations of the singletons boundary fields.
\item{-} Singleton multiplets:

They are just the $N=2$ superconformal multiplets which fall in two categories\foot{Singleton multiplets are listed including antiparticle states.}:
\item{.} Vector multiplets:
\eqn\vec{D(2,1,0|0,0)+D(3/2,1/2,0|2,1)+D(3/2,0,1/2|2,-1)+D(1,0,0|0,2).}
\item{.} Hypermultiplet:
\eqn\hyper{2D(3/2,1/2,0|0,-1)+2D(3/2,0,1/2|0,+1)+D(1,0,0|2,0).}
where the extra two labels are the representations under $SU(2)\times U(1)$.

\item{-} Massless multiplets in $AdS_5$.

Massless multiplets fall in three categories, the graviton multiplet and two
types of matter multiplets: the tensor multiplet and the vector multiplet. As in the maximal case, the tensor multiplet has twice the degrees of freedom of
a vector multiplet. Indeed, in the Poincar\'e limit, after duality, a tensor
multiplet gives two vector multiplets. Note that in $AdS$ the antisymmetric tensor verifies a self-duality constraint in the sense of \warner.

Let us denote by $\Phi, A_i$ (where $i$ is an index in the doublet representation of $SU(2)$) the singleton vector and hypermultiplet,
respectively. $\Phi$ is a multiplet whose first component (lowest $E_0$)
is a complex field, Lie algebra valued in ${\cal G}$, the Yang-Mills group of the world-volume theory. $A_i$  is a multiplet whose first component are two
complex scalars in the doublet representation of $SU(2)$ and in some
representation of ${\cal G}$.

\lref\proyen{M. de Roo, J. W. van Holten, B. de Wit and A. Van Proyen, Nucl. Phys. B173 (1980), 175.}

In absence of hypermultiplets, we may construct the following composite multiplets: $J=Tr\Phi\bar\Phi$ (with maximum spin = 2) and $T=Tr\Phi^2$ (maximum spin = 1). In presence of hypermultiplets, we can construct a third composite, 
$W_l=A_i^aA_j^bt_{ij}\sigma^l_{\alpha\beta}$ ($l=1,2,3$) with value in the adjoint of $G^{\prime}$. These multiplets were listed in \stelle\ and their component expansion given in \proyen\foot{A list of the $AdS_5$ massless multiplets
for various amount of supersymmetry can be found in \warner.}.

\lref\son{M. F. Sohnius, Phys. Lett 81B (1979) 8.}

The $J$ multiplet is the supercurrent multiplet \son, analogous to the $N=4$
counterpart discussed in \ferthree. It contains the graviton, the gravitinos,
the $U(2)$ gauge fields and a $U(2)$ singlet scalar. Its representation content is
\eqn\grav{\eqalign{&D(4,1,1|0,0)+D(7/2,1,1/2|2,-1)+D(7/2,1/2,1|2,1)\cr +&D(3,1/2,1/2|3+1,0) +D(3,1,0|1,-2)+D(3,0,1|1,2)\cr +&D(5/2,1/2,0|2,-1) +D(5/2,0,1/2|2,+1)+D(2,0,0|0,0).}}

The tensor multiplet T is given by the chiral multiplication of
two singleton vector multiplets and its structure is:
\eqn\tens{\eqalign{&D(2,0,0|0,4)+D(5/2,1/2,0|2,3)+D(5/2,0,1/2|2,-3)\cr +&D(3,0,0|3,2)+D(3,1,0|0,2) +D(3,0,1|0,-2)\cr +&D(7/2,1/2,0|2,1)+D(7/2,0,1/2|2,-1)+D(4,0,0|0,0_c).}}

We see that the dilaton scalar, which is massless in $AdS_5$, is its last ($\theta^4$) component.

Finally, if there are hypermultiplets which sustain a flavour symmetry group
$G^{\prime}$, then there is the extra current superfield $W_l$, in the adjoint of $G^{\prime}$, with components:
\eqn\hy{D(2,0,0|3,0)+D(5/2,1/2,0|2,-1)+D(5/2,0,1/2|2,1)+D(3,1/2,1/2|0,0)+D(3,0,0|0,2).}

\lref\wess{ J. Wess and J. Bagger, Princeton Series in Physics, Princeton Univ. Press (1983).}

We now turn to the $N=1$ case. The analysis here is particularly simple because
the $N=1$ superconformal field theories are widely known. We can use here a
superfield notation both for singletons and massless fields in $AdS_5$.

The singleton fields may be described by chiral superfields \wess\ $W_\alpha$, $\phi$,
where $W_\alpha$ is the Lie algebra valued field strenght multiplet of the Yang-Mills singleton boundary theory and $\phi$ is a chiral multiplet in some representation of ${\cal G}$ which makes the theory superconformal invariant \list.

\lref\ferzu{S Ferrara and B. Zumino, Nucl. Phys. B134 (1978) 301.}

The massless composite fields, in absence of the chiral multiplets $\phi$,
are \ferzu: 
\eqn\one{J_{\alpha\dot\alpha}=Tr W_\alpha \bar W_{\dot\alpha},\, D(3,1/2,1/2|0)+D(4,1,1|0)+D(7/2,1,1/2|-3/2)+D(7/2,1/2,1|3/2)}
and
\eqn\two{S=Tr W_\alpha W^\alpha,\, D(3,0,0|3)+D(5/2,1/2,0|3/2)+D(5/2,0,1/2|-3/2)+D(4,0,0|0).}

They correspond to the graviton multiplet and hypermultiplet in $AdS_5$, respectively.

In presence of singleton chiral multiplets, there are two more type of multiplets:
\eqn\three{W=I(\phi\bar\phi ),\, D(2,0,0|0)+D(5/2,1/2,0|-3/2)+D(5/2,0,/1/2|3/2)+D(3,1/2,1/2|0)}
satisfying $\bar D\bar D W=DDW=0$. Here, the symbol I means a singlet under the gauge group ${\cal G}$, and (if there is some $\phi$ in the adjoint of ${\cal G}$)
\eqn\four{\eqalign{T=Tr \phi W_\alpha ,\qquad\qquad &D(5/2,1/2,0|5/2)+D(3,0,0|1)+D(3,1,0|1)\cr +&D(3,0,1|-1)+D(7/2,1/2,0|-1/2).}}

Of course, the $\phi$ multiplets allow extra contribution to eq. \one, and extra hypermultiplets of the type,
\eqn\five{H=I(\phi^2),\qquad D(2,0,0|2)+D(5/2,1/2,0|1/2)+D(5/2,0,1/2|-1/2)+D(3,0,0|-1).}

However, we note that, in both $N=1$ and $N=2$ theories, there is a {\it universal multiplet}, other than the graviton, that contains the type IIB dilaton
and have $E_0=4$; it is massless in $AdS_5$ in the sense that $\partial^2\phi=0$. This universal multiplet is a tensor multiplet ($T$) in $N=4$ and an hypermultiplet in $N=2$ $AdS_5$ supergravity, respectively.

\newsec{$N=2$ and $N=4$ supergravities as Chern-Simons theories.}
A peculiar aspect of odd dimensional supergravity theories is that they require
the presence of Chern-Simons terms, appropriate to the underlying gauge symmetry of the theory.

\lref\senza{M. Gunaydin, G. Sierra and P. K. Townsend, Nucl. Phys B242 (1984) 244.} 

For example, in $D=5,7$, appropriate Chern-Simons terms, $\Omega_5$ and $\Omega_7$, relative to the gauge group $SU(4)$ and $USp(4)$, appear in the maximally extended supergravities in $AdS$ \refs{\guntwo,\pilch}. In the Poincar\'e limit, the (R-) gauge
symmetry becomes abelian and the corresponding Chern-Simons forms just become
the abelian ones \senza.

\lref\sierratwo{G. Sierra, Phys. Lett 157B (1985) 379.}
\lref\sierrathree{M. Gunaydin, G. Sierra and P. K. Townsend, Phys. Rev. Lett. 53 (1984) 322.}

The gauge variation of such terms is a boundary term, which reproduce the 4d anomaly of the global symmetry of the boundary singleton theory. For the $N=8$ case, this was discussed in \witten.

The generic form of the Chern-Simons term in $D=5$ \refs{\guntwo,\awada,\romans,\sierra,\sierratwo,\sierrathree} is
\eqn\chern{\int \Omega_5=d_{\Lambda\Sigma\Delta}\int A^\Lambda\wedge F^\Sigma \wedge F^\Delta.}
where the non-abelian completion is understood in the case where $A$ corresponds to some Yang-Mills symmetry.

In $AdS_5$ with $N=4$ supersymmetry, we know that the pure supergravity part has four vector fields, gauging the R-symmetry group $SU(2)\times U(1)$. Such a
theory has been constructed in \romans. The structure of the Chern-Simons terms in this case is
\eqn\cherntwo{\int B\wedge F^I\wedge F^I.}
where $F^I$ is a $SU(2)$ triplet and $B$ is the $U(1)$ multiplet. Note that no
$\int BF(B)F(B)$ term is present. This means that the theory in $AdS_5$ constructed in \romans\ must correspond
to a superconformal theory on the boundary where the $U(1)^3$ anomaly vanish
but where mixed $SU(2)^2\times U(1)$ anomalies exist and are reproduced by eq.
\cherntwo. Moreover, in this case, no additional continuous symmetries should exist in the singleton theory, otherwise they would reflect in additional vector multiplets
(and Chern-Simons coupling) in the bulk supergravity lagrangian.

Let us now turn to the case where additional flavour symmetries are present.
In this case, following \awada, the Chern-Simons coupling has an additional term $\int B\wedge F^a\wedge F^a$, where $F^a$ are the gauge fields of the flavour symmetry $G^\prime$. In $N=2$ superconformal theories the flavour symmetry
is vector-like. Therefore, the only condition coming from the Chern-Simons terms is still that the anomaly $U(1)^3$ vanishes. It is not known whether $N=4$ supergravities in $AdS_5$ exist with non zero $\int B\wedge F(B)\wedge F(B)$. If
such theory would exist, singleton superconformal field theories with
cubic anomaly $U(1)^3$ would be allowed.

\lref\sagn{A. Sagnotti, Phys. Lett. 294B (1992) 196, hepth/9210127; M. J. Duff, R. Minasian and E. Witten, Nucl.Phys. B465 (1996) 413, hepth/9601036; N. Seiberg and E. Witten,  Nucl.Phys. B471 (1996) 121, hepth/9603003.}

Let us now consider $N=1$ superconformal theories. In this case the Chern-Simons term leaves more freedom, since the constraints on the coefficients $d_{\Lambda\Sigma\Delta}$ are milder \sierra. In this case, the pure $N=2$ supergravity has
a $U(1)^3$ Chern-Simons term, so that the corresponding superconformal theory  
with no flavour symmetry should have a $U(1)^3$ anomaly. It seems to be possible to have many theories. However, it should be pointed out that the $d_{\Lambda\Sigma\Delta}$ enter in the definition of scalar and vector kinetic term, so for particular choices of $d_{\Lambda\Sigma\Delta}$ one may expect to find some singularities for some values of the scalar fields; this is quite analogous to similar phenomena in six dimensional theories with Chern-Simons
terms cancelling gauge anomalies \sagn. The implication of this would result in a loss of validity of the supergravity lagrangian in $AdS_5$ to describe 4d superconformal field theories.
\newsec{Some candidate dual pairs of superconformal and supergravity theories.}
We can use the correspondence between Chern-Simons terms in supergravities in
$AdS_5$ and anomalies of global symmetries in the superconformal theories, as
a guideline for finding dual pairs of theories or checking proposed ones. 

In
\refs{\silver,\vafa}, superconformal theories with $N=1,2$ supersymmetry, which should have a supergravity dual in $AdS_5$, were
found by orbifoldizing the original example with $N=4$ supersymmetry in \malda.
In general, we expect to find candidates for theories with a supergravity dual
in the class of finite supersymmetric gauge theories. In the $N=2$ case,
the vanishing of the one-loop beta function is enough to guaranteer the finiteness of the theory, while in the $N=1$ case higher loops must be checked.
A list of finite $N=2$ and $N=1$ supersymmetric gauge theories can be found
in \list. 

Let us start with the $N=2$ case and the pure supergravity in $AdS_5$ constructed in \romans.  The theory in \romans\ has a gauge group $U(1)\times SU(2)$ which corresponds to the R-symmetry of the superconformal boundary theory. The constraint coming from the supergravity Chern-Simons
terms is that the $U(1)^3$ anomaly must vanish while the $SU(2)^2\times U(1)$ anomaly must be generically non zero. 
An example of a $N=2$ superconformal theory,
which should admit a $AdS_5$ description, was described in \refs{\silver}, by
orbifoldizing the $N=4$ case in \malda. The theory has Yang-Mills group $U(n)^k$ and matter in the
\eqn\matter{(n,\bar n,1,1,...)+(1,n,\bar n,1,...)+...(\bar n,1,...,n).}
and it is the world-volume theory for D3 branes sitting at an orbifold singularity.
The $U(1)^3$ anomaly is indeed zero, while the $SU(2)^2\times U(1)$ anomaly, which does not receive contributions from the hypermultiplets, is
different from zero. The fact that the decoupled $U(1)$ factor in the superconformal theory must be taken into account in order to cancel the $U(1)^3$
anomaly, is a signal that the supergravity theory describes also the center-of-mass motion of the D3 branes. Note that we cannot exclude the presence of additional matter multiplets in the supergravity theory.

Let us discuss the case in which there are additional flavours symmetries. A standard example of $N=2$ superconformal theory is $SU(n)$ with $2n$ flavours. The  $U(1)^3$ anomaly does not vanish, indicating that this theory cannot have a supergravity dual in the class of theories constructed in \refs{\romans,\awada}. The existence of different $N=4$ supergravity theories in $AdS_5$ which
allow a cubic Chern-Simons term for $U(1)$ cannot be excluded a priori. In any case, since the duality proposed in \malda\ is
valid only for large n, a candidate supergravity dual would have an infinite
number of vector multiplets corresponding to the $SU(2n)$ flavour symmetry,
thus reducing the predictivity of such a duality. Superconformal theories with
flavour symmetry which does not increase with $n$ can be found in \list. Their
supergravity dual must be searched in the class of theories discussed in \awada\ (provided that the cubic $U(1)_R$ anomaly vanishes).

\lref\flato{M. Flato and C. Fronsdal, {\it Interacting Singletons}, hepth/9803013.}

Let us now consider the $N=1$ case. The anomaly constraints are now far less
restrictive. In general, the Chern-Simons terms in the $N=2$ pure supergravity in $AdS_5$, which is common to all the models, implies the existence of a $U(1)^3$ R-symmetry anomaly in the
superconformal theory on the boundary. In the presence of other gauge fields in
$AdS_5$, corresponding to global symmetries on the boundary, the general
form of the coefficients $d_{\Lambda\Sigma\Delta}$ gives the form of the expected anomalies. The identification of the global symmetry and the matching of the anomalies is, in general, not enough for identifying dual theories, but gives
strong constraints on the candidate pairs.

Let us consider a particular example. In \sierra, a particular $N=2$ supergravity in $AdS_5$ with
gauge group $U(1)\times SU(3)$ was constructed. This theory contains eight vector multiplets in $AdS_5$, whose scalar partner parametrise the exceptional manifold $SL(3,C)/SU(3)$. The form of the coefficients,
\eqn\coeff{d_{000}=1, d_{00a}=0, d_{0ab}=-{1\over 2}\delta_{ab}, d_{abc}=D_{abc}.}
where the indices 0 and a label $U(1)$ and $SU(3)$ respectively, and $D_{abc}$ is the symmetric tensor of $SU(3)$, suggests the existence of $SU(3)^3$ and $SU(3)^2\times U(1)$ anomalies. The dual theory must be a $N=1$ superconformal theory which has a $U(1)_R\times SU(3)$ global symmetry with the same anomalies.
A candidate superconformal theory with a  $U(1)_R\times SU(3)$ global symmetry and the same anomalies was discussed in \silver. It has a Yang-Mills group $U(n)^3$ and matter consisting of chiral multiplets in the representation,
\eqn\enne{3((n,\bar n,1)+(1,n,\bar n)+(\bar n,1,n))}
Since it is constructed as an orbifold of the original D3 branes theory in \malda, we expect that a supergravity description in $AdS_5$ should exist. Such a description could be provided by the $N=2$ supergravity constructed in \sierra, with possible additional matter multiplets. 
\newsec{Conclusions}
In this paper we have studied the general structure of $N=2$ and $N=4$ supergravities in $AdS_5$, with massless multiplets as {\it composite operators} of
the boundary singleton theory \refs{\pol,\fertwo,\witten,\flato}.

In analogy to the $N=4$ case, we have associated the supermultiplets with $J_{MAX}=2$ to the supercurrents multiplet which always contains the graviton and the gauge bosons of the R-symmetry group ($U(2)$ for $N=2$ and $U(1)$ for $N=1$).
Unlike the $N=4$ theories, other multiplets exist for $N=2,1$ as composite boundary excitations. 

For $N=4$ supergravities, we have vector and tensor multiplets (both with $J_{MAX}=1$). The former are related to additional flavour symmetries (carried by the boundary hypermultiplets) of the $N=2$ superconformal field theory. A universal tensor multiplet is related to the dilaton.

For $N=2$ theories in $AdS_5$, we may have, other than the graviton multiplet, three types of matter composite multiplets: vectors (composite of boundary multiplets carrying some flavour symmetry),
tensors and hypermultiplets. The bulk hypermultiplets are related to bound states of chiral superfields. One of them is the universal bulk dilaton hypermultiplet.

An important constraint for the study of $N=1,2$ superconformal field theories, which have a supergravity interpretation in $AdS_5$, in the sense of \refs{\malda,\pol,\witten}, is that the global anomalies on the D3 world-volume must be reproduced by the 5D Chern-Simons couplings of the corresponding supergravity.
This gives a remarkable constraint. For example, the $N=4$ supergravity of
refs. \refs{\romans,\awada} does not allow a $U(1)^3$ R-symmetry anomaly. This seems to imply that only boundary singleton theories without such an anomaly can be associated to such theory. We do not know whether a $N=4$ theory with a 
$U(1)^3$ invariant may be constructed, allowing to reproduce more models.

In the case of $N=2$ supergravity in $AdS_5$, the Chern-Simons term allows a much richer structure. In fact, if the corresponding $N=1$ superconformal boundary theory does not carry flavour symmetries, there is a unique $U(1)^3$ anomaly whose Chern-Simons term is present in the $N=2$ theory on $AdS_5$.

An interesting question is to understand how many composite tensor and hypermultiplets are present in $AdS_5$ for a given superconformal singleton theory on the boundary. It is perhaps possible to give a simple answer also to this question.

The presence of a scalar potential for supergravities in $AdS_5$ allows to study critical points for different possible vacua in the bulk theory (a general analysis was given in \warner. It is natural to conjecture that these critical points should have a dual interpretation in the boundary superconformal field theory side.

The increasing evidence of a correspondence between supergravity theories
may very well go beyond the original interpretation of singletons as branes degrees of freedom and $AdS_5$ as nearly horizon geometry of the brane. It could
in fact, as also discussed in \refs{\pol,\witten,\sjr,\maldathree,\arefa}, reveal a powerful tool
in the study of non-perturbative aspects of conformal invariant quantum field
theories.
 
\centerline{\bf Acknowledgements}
S. F. is supported in part by DOE under grant DE-FG03-91ER40662, Task C, and by ECC Science
Program SCI*-CI92-0789 (INFN-Frascati)
\listrefs
\end